# What Makes You Hold on to That Old Car? Joint Insights from Machine Learning and Multinomial Logit on Vehicle-level Transaction Decisions

**Ling Jin[1], Alina Lazar[2], Caitlin Brown[1], Bingrong Sun[3], Venu Garikapati[3], Srinath Ravulaparthy[1], Qianmiao Chen[1], Alexander Sim[1], Kesheng Wu[1], Tin. Ho[1], Thomas Wenzel[1], C. Anna Spurlock[1*]**
[1]Lawrence Berkeley National Laboratory, Berkeley, CA 94720
[2]Youngstown State University, Youngstown, OH 44555
[3]National Renewable Energy Laboratory, 15013 Denver West Parkway, Golden, Colorado 80401

*** Correspondence:**
Corresponding Author
caspurlock@lbl.gov

**Abstract**
What makes you hold on that old car? While the vast majority of the household vehicles are still powered by conventional internal combustion engines, the progress of adopting emerging vehicle technologies will critically depend on how soon the existing vehicles are transacted out of the household fleet. Leveraging a nationally representative longitudinal data set, the Panel Study of Income Dynamics, this study examines how household decisions to dispose of or replace a given vehicle are: (1) influenced by the vehicle's attributes, (2) mediated by households' concurrent socio-demographic and economic attributes, and (3) triggered by key life cycle events. Coupled with a newly developed machine learning interpretation tool, TreeExplainer, we demonstrate an innovative use of machine learning models to augment traditional logit modeling to both generate behavioral insights and improve model performance. We find the two gradient-boosting-based methods, CatBoost and LightGBM, are the best performing machine learning models for this problem. The multinomial logistic model can achieve similar performance levels after its model specification is informed by TreeExplainer. Both machine learning and multinomial logit models suggest that while older vehicles are more likely to be disposed of or replaced than newer ones, such probability decreases as the vehicles serve the family longer. Pickup trucks and sport utility vehicles are less likely to be disposed of or replaced than cars, and leased vehicles are more likely to be transacted than owned vehicles. We find that married families, families with higher education levels, homeowners, and older families tend to keep their vehicles longer. Life events such as childbirth, residential relocation, and change of household composition and income are found to increase vehicle disposal and/or replacement. We provide additional insights on the timing of vehicle replacement or disposal, in particular, the presence of children and childbirth events are more strongly associated with vehicle replacement among younger parents.

## 1 Introduction

Previous studies have shown that the choice of whether or not to own a vehicle, and, if so, what type(s), is a medium-term decision that is shaped by life contexts. For example, vehicle ownership is influenced by household socio-demographic (household size) and economic (income) characteristics, proximity of home location to work and other locations, and life-stage transitions, such as the birth of a child [1] and changes in the number of adults in the household [2]. Vehicle transaction decisions (add, replace, or dispose of vehicles) take place at different stages along a household's life-course and co-evolve with changes in residential and workplace location [3].

Due to the paucity of longitudinal data on vehicle transactions, most existing literature relies on cross-sectional data and provides only a static snapshot of vehicle ownership as reviewed in [4], [5] However, research on mobility biography and life-oriented approaches [1], [6]–[9] has long recognized the interdependence of choices across various life-stages and recommends integrating intertemporal dynamics into the analyses of long-term mobility in a comprehensive way. An increasing number of studies have followed the mobility biography approach to understand vehicle ownership evolution dynamics over a given household's lifetime [1], [2], [7], [10].

Vehicle transaction dynamics investigated by previous studies include changes in the vehicle holdings (e.g. [10], [11]) and vehicle transaction decisions [2], [12], [13][14]. However, changes in the vehicle holdings cannot uncover the incidence of vehicle replacement, as it does not alter the level of vehicle holdings. At the same time, most of the dynamic vehicle ownership studies focus on household-level decisions without including vehicle attributes nor further determining which vehicle will be disposed or replaced. This omission limits the applicability of these vehicle ownership models, as vehicle-level transaction decisions are needed for forecasting fleet dynamics over a 10- to 20-year horizon, such as is needed in microsimulations to project fleet evolution in the case of transportation decarbonization policies. While the vast majority of light-duty vehicles are still powered by ICEs [15], adoption of emerging vehicle technologies, such as those with electric drivetrains, will depend on how quickly ICE vehicles are transacted out of household fleets.

When it comes to prediction methods, logit models have long been the gold standard in choice modeling for transportation behavior (see reviews [4] [16][17]). These choice models are based on random utility maximization theory, and the estimated coefficients can be readily interpreted as changes in odds ratios. Unlike statistical models that impose a predetermined structure, machine learning (ML) models on the other hand rely on data-driven heuristics to arrive at their solutions. In recent years, ML methods have been adopted for travel behavior studies, including mode choice [18]–[20][21], route choice [22], [23], activity type choice [24], [25] and joint decisions such as departure time and mode choice [26], [27].

Despite the broad application of ML in travel behavior modeling, limited attention has been paid to using ML methods in vehicle transaction modeling. Predicting the dynamics of vehicle transactions requires longitudinal data that are difficult to collect from the life courses of individual households. Current data collection is mostly reliant on cross-sectional surveys with small sample sizes that limit the application of data-driven ML models. Furthermore, while ML has advantages with respect to handling large datasets efficiently, limitations in interpretability of ML results constrains their applicability to behavioral research use cases in which the goal is to better understand and design transportation policies [17].

Rather than treating the gold standard and ML models as competitors, opportunities exist to marry the two. Recent advances in "Explainable AI" [28] have improved the interpretability of tree-based ML models exploring high-dimensional feature spaces. Behavioral insights from ML models, such as individual feature importance, directional influences, and feature interactions, have been incorporated into the logistic model building process to improve model specification and prediction performance [29], [30]. Travel behavior insights jointly determined from ML and logistic models can also improve the robustness of derived conclusions.

This study seeks to fill two research gaps in the literature: (1) the need for capturing fleet dynamics accurately by addressing vehicle-level disposal and replacement decisions; and (2) inadequate application of both choice modeling and ML methods to vehicle transaction dynamics due to limited longitudinal data collected over time. Leveraging a nationally representative panel data set, the Panel Study of Income Dynamics (PSID), we examine how the likelihood of disposing or replacing a vehicle is: (1) influenced by vehicle attributes, (2) mediated by concurrent family socio-demographic and economic status, and (3) triggered by key life cycle events. While past studies have separately investigated one or two of the above dimensions, this paper is the first to relate all three simultaneously to vehicle-level disposal and replacement decisions. Additionally, we advance the empirical analysis methodology by coupling machine learning models with a recently developed TreeExplainer [28] as an additional interpretation tool to both generate behavioral insights and improve the model specification for choice modeling.

Our contributions to travel behavior research include the following: (1) ours is the first study to model vehicle-level transactions using revealed choices in a national panel survey in the U.S.; (2) our approach seeks to simultaneously investigate the effects of vehicle attributes and household concurrent and life event attributes on vehicle level transaction decisions; and (3) we demonstrate an innovative use of ML methods to inform the model building process for choice modeling and to jointly generate behavioral insights and improve model performance. The effects of various predictor attributes derived from both ML and logit modeling are compared qualitatively to improve the robustness of our results. Policy insights are also discussed in relation to our findings.

## 2 Materials and Methods

### 2.1 Data Source Preprocessing and Description

To understand vehicle transaction dynamics, time varying attributes are needed from longitudinal data, including vehicle attributes, household characteristics, and life-stage events. We employ nine biennial waves, 2003 through 2019, from the publicly available version of the Panel Study of Income Dynamics (PSID) [31], the longest-running national-level longitudinal panel survey of American families. Since 1968, PSID collected data from a sample of U.S. families over time primarily focused on questions pertaining to family expenditures and income. PSID also has a wealth of information on the social, economic, and demographic characteristics of individuals and families. Due to its panel structure and long history, PSID data has become an important data source for life course research [10], [32]–[34]. PSID initially collected limited vehicle ownership information, such as the number of vehicles included in each family. Since 1999, PSID has collected individual vehicle information for up to three vehicles in each family, that together covers 95% of the total number of vehicles reported by the families. The captured vehicle information includes body type, model year, whether the vehicle is owned or leased, acquisition year, manufacturer, and make. We limit our study to the PSID survey waves from 2003 to 2017, because they include consistent questions about vehicle information, and life event variables can be consistently derived between survey waves. Due to the biennial nature of the survey, the exact timing of life events and vehicle transactions are subject to some uncertainties.

### 2.1.1 Preprocessing Transaction Outcomes of Existing Vehicles

The outcome variable of interest in this study is the transaction decision for individual vehicles in the family's existing fleet. That is, for any given vehicle of the current survey wave, we predict whether it is disposed of without replacement ("disposed" hereafter), disposed of with replacement ("replaced" hereafter), or kept in the family ("kept" hereafter) by the next survey wave. In contrast from most existing literature, where vehicle transaction decisions were readily reported in stated-preference or retrospective surveys [2], [13], [35], this outcome variable needs to be generated by tracking the revealed vehicle information from one wave to the next in the PSID.

Vehicles reported in the PSID do not include unique identifiers, and hence it is not straightforward to determine the presence or absence of a given vehicle in consecutive waves. We first create a unique identifier for each vehicle (vehicle id) reported in each wave using a combination of the unique person id of the head-of-household, vehicle model year, manufacturer, make, and vehicle body type. These derived identifiers uniquely identify 99.8% of the vehicles in PSID survey waves used in this study. The presence of the same vehicle ids from wave to wave represents a given vehicle's life trajectory in a household (as defined by the household head) over time. From the life trajectory of individual vehicles, we then determine the timing when a that vehicle is added to the household fleet and when it is removed from the fleet. Finally, the transaction outcome ∈{disposed, replaced, kept}of an existing vehicle in a given wave can be determined by its presence or absence in the next wave, in conjunction with whether vehicle acquisition is observed. Specifically, in the case when a vehicle was removed from the household in the next wave, its outcome is coded as "replaced" when the family concurrently adds another vehicle to their fleet during the two-year window; whereas in the case of no vehicle acquisition observed, the outcome of the removed vehicle is coded as "disposed of" with no replacement.

Over the nine waves, on average 10% of the vehicles in each wave are disposed of before the next wave, 31% are replaced, and the remaining 59% are kept (**Table 1**). The distribution of vehicle outcomes is relatively stable over the years considered in the data (**Figure 1 a**), except that the next-wave disposal rates are slightly higher and replacement rates are lower for vehicles observed during the 2007 - 2011 waves, which may be due to the effects of the 2008 economic recession.

### 2.1.2 Description of Explanatory Variables

Both vehicle-specific and the family level attributes are considered to explain the transaction outcome for individual vehicles.

The vehicle-specific attributes derived from the PSID survey include vehicle model year, number of years serving the family (i.e., years since it was first acquired by the family), whether the vehicle is owned or leased, and vehicle body type. Vehicles observed in each wave are on average 9.6 years old and have served their respective households for about 6 years (**Table 1**). The vintage composition experienced a shift after the 2009 survey wave, with younger vehicles (<=5 years) decreasing from 45% in 2007 to the lowest point of 30% in 2013, while older vehicles (>=12 years) increased from 16-18% in 2007-2009 to 25% in 2017 (**Figure 1 b**). This trend is consistent with the nation-wide increase in older vehicles in U.S. households revealed in NHTS 2009, and 2017 snapshots [36]. Most of the vehicles (54%) are of body type 'car' (**Table 1**), though the share of vehicles with this body type has decreased over the years (from 58% in 2003 to 50% in 2017) (**Figure 1c**). In contrast, shares of SUVs have increased steadily from 16% in 2003 to 30% in 2017. While the majority of the vehicles are owned rather than leased (**Table 1**), the percentage of vehicles leased has doubled from ~3-4% before 2011 to ~8% in 2017 (**Figure 1d**).

These vehicle attributes vary across outcome categories (**Table 1** last three columns). Disposed vehicles tend to be the oldest among all categories. On average, disposed vehicles are 1-2 years older than replaced vehicles, and 2-3 years older than vehicles that are kept through to the next survey wave. Distribution of body types are also different among the vehicle transaction categories. For example, a greater proportion of disposed vehicles belong to the ‘car’ body type, and pickups make up the majority category in the vehicles retained by households. The association of vehicle attributes with observed vehicle transaction outcomes will be determined more quantitatively in the empirical analysis section.

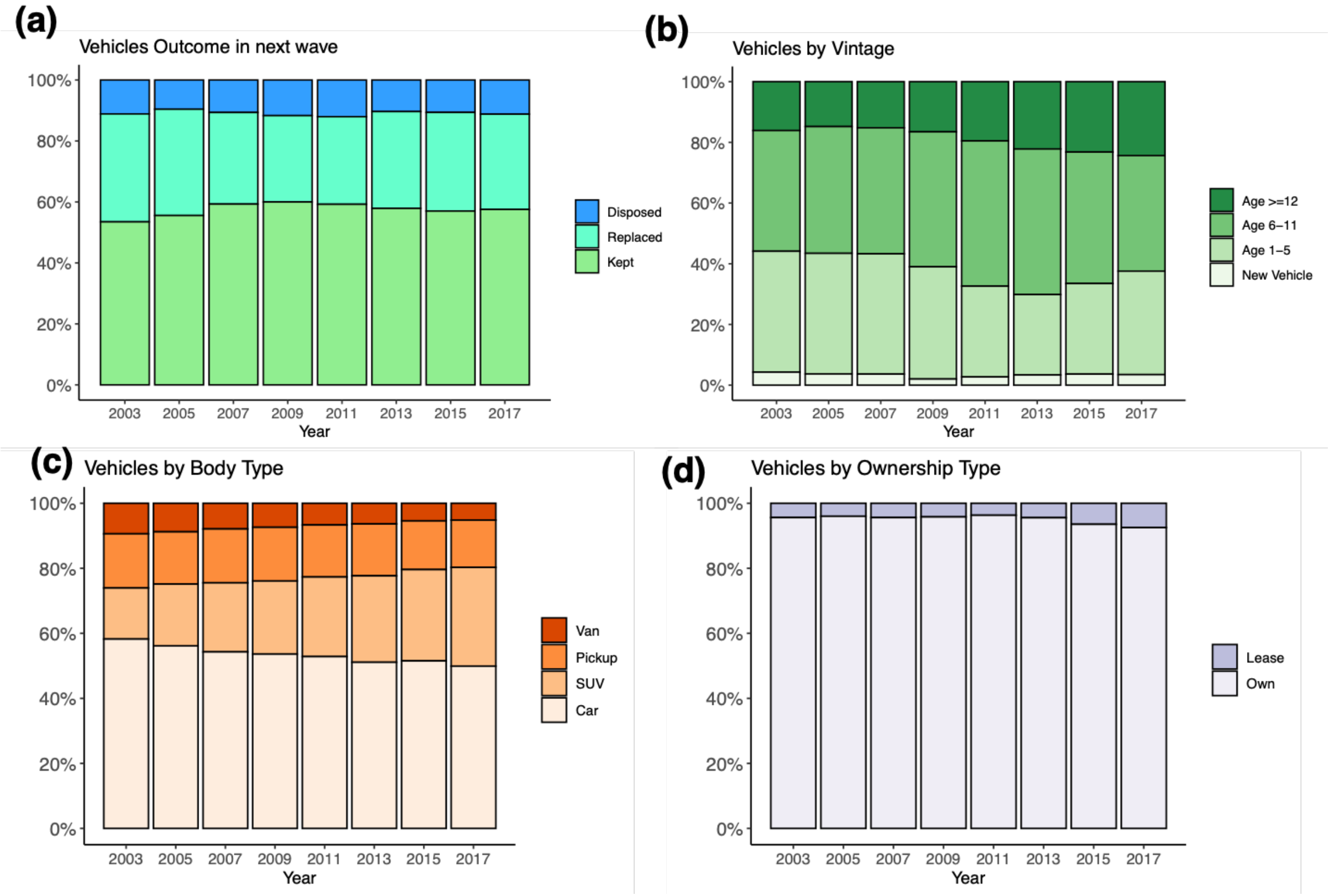

**Figure 1. Summary of distribution of vehicle variables by year: (a) vehicle outcome; (b) vehicle vintage; (c) vehicle body type; (d) vehicle ownership type.**

Family-level attributes are processed to develop both static socioeconomic characteristics of the family at the concurrent wave of the vehicle and “change” variables that represent the life events occurring between the current and next wave. Current wave attributes include household fleet size, number of eligible drivers (>= 16 years old), number of children and presence of children by their age bins, number of workers, income levels, marriage/cohabitation status, education level, and built environment characteristics such as house type and tenure. The population means for these attributes are presented in **Table 1**. Access to restricted PSID location data at census tract or block level is required to derive additional built environment attributes pertaining to the home and work locations of households. Inclusion of this information is left to future work.

Change variables are generated from family and individual socio-demographic time varying attributes as well as life events explicitly surveyed in the PSID, with variable selection largely inspired by the existing mobility biography literature [10], [37]. These include change in income, increase in head or spouse’s education level, change in number of workers and drivers, and additional life events such as child birth, cohabitation, marriage, divorce, retirement or death of a family member, residential relocation, and empty nesting (i.e., all grown up children moving out of the household). The average occurrence rates of familial events are found to be less than 10% (Population mean column in **Table 1**). Residential relocation is the most frequent of all included life events, which happened to 45% of the families from wave to wave.

Similar to the vehicle attributes, many of the family level attributes vary across vehicle transaction outcomes as indicated by the descriptive statistics (**Table 1** last three columns). For example, disposed vehicles seem to be associated with families with more than 2 vehicles (disposed vehicle included), while replaced or kept vehicles are associated with families with less than 2 vehicles available. In addition, disposed vehicles are associated less with home owners, and are associated more with low-income and low education-level families, and families with steeper decreases in income. Replaced vehicles are more frequently associated with presence of young children. A more quantitative understanding of these family level mediating factors and life-event triggers for the observed vehicle outcomes will be described in the empirical analysis results.

**TABLE 1 Descriptive Statistics of Vehicle Specific and Family level Variables for the Full Sample and by Vehicle Outcome.**

| **Vehicle-level Summary**<br>Variable Description (short name) | **Population Mean** | **By Vehicle Outcome in the Next Wave** | | |
|---|---|---|---|---|
| | | **Kept** | **Disposed** | **Replaced** |
| Vehicle vintage (vintage) | 9.55 | 8.95 | 11.63 | 9.98 |
| Years in family (yrs_inFu) | 6.06 | 6.25 | 6.12 | 5.70 |
| Owned (ownlease) | 0.95 | 0.97 | 0.95 | 0.91 |
| Vehicle body type (vehtype) | | | | |
| Car | 0.54 | 0.52 | 0.59 | 0.55 |
| Pickup | 0.15 | 0.16 | 0.14 | 0.13 |
| Utility | 0.24 | 0.25 | 0.2 | 0.24 |
| Van | 0.07 | 0.07 | 0.07 | 0.08 |
| Vehicle outcome in the next wave | | | | |
| Disposed (w/o replacement) | 0.1 | 0 | 1 | 0 |
| Kept | 0.59 | 1 | 0 | 0 |
| Replaced | 0.31 | 0 | 0 | 1 |
| Number of observations (vehicle-year) | 69,697 | 40,884 | 7,178 | 21,635 |
| **Family-level Summary**<br>Variable Description (short name) | **Population Mean** | **By Vehicle Outcome in the Next Wave** | | |
| | | **Kept** | **Disposed** | **Replaced** |
| ***Current wave status*** | | | | |
| Number of vehicles (Nveh) | 1.98 | 1.94 | 2.3 | 1.95 |
| Age of family head or spouse (hh_age) | 46.05 | 47.59 | 46.16 | 44.07 |
| Number of drivers (Ndrivers) | 1.94 | 1.99 | 2.18 | 2.05 |
| Number of children (Nkids) | 0.73 | 0.68 | 0.73 | 0.86 |
| Presence of children <=4 yrs old (kid_4) | 17% | 16% | 18% | 21% |
| Presence of children 5-11 yrs old (kid_5_11) | 24% | 22% | 23% | 27% |
| Presence of children 12-15 yrs old (kid_12_15) | 14% | 14% | 15% | 16% |
| Number of employed (N_emp) | 1.38 | 1.35 | 1.37 | 1.43 |
| Income (inc_5bins) | | | | |
| < $25,000 | 11% | 9% | 17% | 11% |
| $25,000 – $50,000 | 21% | 20% | 23% | 20% |
| $50,000 – $75,000 | 20% | 20% | 19% | 20% |
| $75,000 - $150,000 | 34% | 35% | 28% | 34% |
| >= $150,000 | 15% | 15% | 13% | 15% |
| Education level (edu) | | | | |
| Less than high school | 5% | 4% | 9% | 5% |
| High school | 26% | 24% | 31% | 26% |
| Some college | 29% | 28% | 31% | 30% |
| College degree | 21% | 22% | 16% | 20% |
| Post-graduate | 19% | 21% | 13% | 18% |
| Married or cohabiting (spouse) | 65% | 66% | 58% | 67% |
| Home owner (house_tenure) | 65% | 69% | 56% | 62% |
| Live in a house (house_type) | 84% | 85% | 81% | 83% |
| ***Life event and change variables from current to next wave*** | | | | |
| Change in income (in thousand) (ch_income) | 1.26 | 1.28 | -3.98 | 2.99 |
| Increase edu levels (ch_Edu) | 0.05 | 0.05 | 0.05 | 0.06 |
| Child birth indicator (birth) | 0.1 | 0.09 | 0.09 | 0.12 |
| Empty nest (kid_moveout) | 0.07 | 0.05 | 0.16 | 0.06 |
| Family member retired indicator (retire) | 0.06 | 0.06 | 0.06 | 0.05 |
| Family member died indicator (death) | 0.01 | 0.01 | 0.01 | 0.01 |
| Change of number of employed (ch_Nemp) | -0.05 | -0.03 | -0.23 | -0.03 |
| Change of number of drivers (ch_Ndriver) | -0.02 | 0.01 | -0.26 | 0.005 |
| Family moved recently (moved) | 44% | 39% | 50% | 49% |
| Change of family head marriage (ch_marriage) | | | | |
| No change | 96% | 97% | 91% | 95% |
| Coupling up | 2% | 2% | 1% | 3% |
| De-coupling | 2% | 1% | 7% | 2% |
| Number of observations (family-year) | 42,052 | 29,779 | 6,221 | 18,628 |

## 2.2 Empirical Analysis and Interpretation Approach

We employ both machine learning and multinomial choice modeling for empirical analysis. However, unlike prior studies that pitted the ML models against traditional choice models, in this study we utilize the patterns and insights learned from ML models to inform specifications of the choice models, such as justifying the binning of the continuous variables and determining the interaction terms. We show that this ML-aided model building process improves the performance of the multinomial logit model.

The unit of analysis is household vehicles and therefore the outcome variable is at the vehicle level: for each vehicle observed in a given survey wave, the models predict its transaction outcome ∈{disposed, replaced, kept}in the next wave. Input features (i.e. independent variables) include vehicle attributes, current fleet size, socio-demographic attributes of the family in the current wave, and the life event change variables determined from current to next wave.

### 2.2.1 Machine Learning Method and TreeExplainer

Application of machine learning to predicting vehicle transactions is new. Existing literature has focused on travel mode predictions (see review by [16]) and there has been no documentation on both the performance of various ML methods on predicting vehicle transactions and their comparison to the gold standard logit models. Four ML methods are first evaluated by this study and the best performing method is then coupled with TreeExplainer to further generate behavioral insights and inform model specification in logit modeling. The four algorithms evaluated are:

- *Random Forest.* This algorithm [38] builds an ensemble of decision trees, or tree predictors, which depend on randomly and independently sampled vectors over the same distribution. The strength, correlation and monitor error are closely followed to track the growing features in response to the branches splitting.
- *CatBoost and LightGBM.* Standard gradient boosting methods are based on random forest, aiming to solve over-fitting problems, but inefficiently. In an effort to make gradient tree boosting more flexible and scalable, Chen [39] created the scalable XGBoost algorithm. XGBoost employs a regularization technique to minimize over-fitting. This tactic allows XGBoost to be faster and more robust during tuning. Because the majority of input features are categorical variables, we employ the two gradient boosting based methods, CatBoost and LightGBM, that were shown to have better performance for categorical data [40]. Both these methods are extensions of XGBoost. CatBoost focuses on categorical columns using permutation techniques and target-based statistics [41]. The light gradient boosting machine (LightGBM) further improves standard gradient boosting methods. Microsoft developed LightGBM by growing the decision trees leaf-wise, allowing it to effectively utilize GPU for faster training time and better accuracy [42].

- *Neural Network - Multilayer Perceptron.* One of the simplest multi-layered neural network architectures, the multilayer perceptron (MLP) [43] is a hierarchical structure of layers containing individual artificial neurons. The power of MLPs comes from their ability to learn patterns in the training data and to relate them to the output. Mathematically, MLPs are considered universal approximators, which means they are capable of learning any mapping function. The MLP architecture consists of an input layer, one or more hidden layers, and an output layer. Each neuron in the hidden layer receives input from the preceding layer and fires according to the neuron's activation function. During the forward pass, the output of each layer is passed to the next layer and the output layer consists of only one neuron. The error is calculated based on the function to be predicted and the output of the network. After the forward pass, the backpropagation algorithm [44] is used to adjust the model's weights and biases. This combination of forward and backward passes is repeated for many epochs until some stopping criterion is satisfied. This whole process is called training. After training, the resulting model can be used for classification and prediction.

For application in policy and behavioral analyses, explaining and interpreting the predictions made by these machine learning models is critical, but not trivial. The most exciting recent development in explaining tree-based methods is the "TreeExplainer" by Lundberg et al. [28]. Most of the existing machine learning studies interpret variables by their global importance ranking (e.g., using Gini index [29], [45]) which may mask their local importance when interacting with other variables. Additionally, Gini coefficient does not provide any indication of the direction of association, which is critical in interpreting the results of a model for policy evaluation purposes. The TreeExplainer fills this gap by computing the optimal local explanations for the variables including both the sign and magnitude of their effects.

The key quantity computed by the TreeExplainer is Shapley Additive exPlanation (SHAP) values, which represents the sequential impact on the model's output of observing each input feature, averaged over all possible subset variable orderings.

$$\phi_{jntl} = \sum_{S\subseteq N\setminus\{l\}} \frac{|S|!(M-|S|-1)!}{M!}\left[f_x(S\cup\{l\}) - f_x(S)\right] \quad (1)$$

Where $\phi_{jntl}$ is the SHAP value of the $l$-th variable on the outcome class $j$ for vehicle $n$ at wave $t$. S represents a subset of variables that do not include the $l$-th variable, |S| is the total number of variables in the subset, and M is the total number of variables. $f_x$ is the prediction function.

Aside from providing the both local and global ranking of the input variables based on their contribution to the classification, the SHAP values could also be used to plot individualized explanations for each feature and their localized effects on the final prediction. The SHAP dependence plots provide richer information than traditional partial dependence plots [28]. These plots capture both the direction and the magnitude of impact of each variable as well as the interaction between variables on the classification task. In addition to the visual guide provided by SHAP dependence plot, SHAP methodology allows for quantitative determination of salient interactions of tree-based models by expanding the method to include interaction terms for individual observations [28]. A measure of the global importance of these interactions can be characterized by summing the interaction effects over all of the samples to find the important interaction terms to add as predictors to the choice modeling.

### 2.2.2 Multinomial Choice Modeling and Interpretation

We apply the multinomial logit (MNL) model to predict the transaction outcome of each individual vehicle in the dataset. MNL models are the most widely used choice models, and are based on the principle of random utility maximization derived from econometric theory. The utility of keeping the vehicle in the next survey wave is fixed at 0 without any loss of generality, while the utility function from choosing alternative $j$, that is, to dispose or replace, the vehicle $n$ of wave $t$ in family $i$ is defined as:

$$U_{njit} = \alpha_j + X'_{nt}\beta_j + Z'_{it}\gamma_j + Year_t \cdot \delta_j + \varepsilon_{njit} \qquad (2)$$

$\beta_j$ is the alternative specific coefficient vector associated with the vehicle attributes $X'_{nt}$, and $\gamma_j$ is the alternative specific coefficient vector associated with the family level socio-demographics attributes $Z'_{it}$. To account for the temporal trends discussed in Section 2.1, we include year-specific effects $Year_t \cdot \delta_j$ in our model. To account for serial correlation across time observations within families, we cluster the standard errors of the estimates at the level of the household.

The baseline MNL (bMNL) model is built with all the input features entering the equation linearly. The final model specification, referred to as improved MNL (iMNL), will be informed by the TreeExplainer coupled with the best performing ML model, such as decisions about binning of continuous variables and whether and how to include interaction terms. The baseline MNL model is intended to establish a performance baseline, from which the improved MNL model can be compared relative to the ML models.

Note that we did not choose a nested structure where households first decide whether to do nothing, dispose, add, or replace a vehicle, then in the nested layer determine which vehicle to dispose or replace. Such nested structures are not viable owing to the constraints presented by the PSID data where family level decisions revealed during the two-year window are sometimes not mutually exclusive.

The interpretation of the coefficients in MNL models is straightforward by examining the sign and significance level of the coefficients. For a unit change in the predictor variable, the utility of vehicle transaction outcome $j \in$ {disposed, replaced} relative to the "kept" decision is expected to change by its respective coefficient estimate, given that the other variables in the model are held constant. Therefore, a positive value of the coefficient means that the vehicle is more likely to be replaced or disposed, relative to being kept. The sign of the coefficients will be compared with the TreeExplainer results to derive behavioral insights more robustly.

Note that traditional variable selection employs LASSO or Ridge regression techniques to penalize the model's complexity in the presence of a large number of predictors. While this study does not directly employ LASSO for variable selection, we use it for confirming the validity of the variable selected by ML models coupled with the TreeExplainer.

### 2.2.3 Performance Evaluation Methods

Ten metrics are used to comprehensively evaluate various aspects of the performance of both ML and MNL models for predicting multi-class vehicle transaction outcomes.

The outcome class-specific metrics such as Accuracy, Recall, F1, and Specificity are first computed from the confusion matrices. Then the following multi-class overall performance metrics are derived:

- *Overall Accuracy for correct classification*, which indicates the fraction of instances that are correctly classified.
- *Average Accuracy*, which is based on the sum of the one-vs-all matrices, and represents a binary classification task where one class is considered the positive class and the combination of all the other classes make up the negative class.
- *Macro-averaged metrics*, which include Macro-precision and Macro-F1, also known as sensitivity or the true positive rate, is calculated by taking the means of per-class precision, recall and F1, respectively.
- *Micro-averaged metrics*, which are from the sum of the one-vs-all matrices for each class, and the sum of these matrices will always be a symmetric matrix, so micro-precision, -recall and -F1 will be the same.

Three additional overall performance metrics that do not rely on the class-specific metrics are also computed including:

- *Cohen's Kappa*, which can be interpreted as a comparison of the overall accuracy to the expected random chance accuracy with higher value indicating a better classifier compared relative to a random chance classifier.
- *Cross-entropy*, which measures the difference between two probability distributions from the idea of entropy in information theory to quantify the number of bits required to transmit an average event from one distribution to another distribution. Lower cross-entropy indicates better model performances.
- *Multi-class Log Loss*, which penalizes the model for uncertainty in correct predictions, and heavily penalizes the model for making an incorrect prediction, with lower multi-class log loss as better model performances.

# 3 Results

## 3.1 Machine Learning Model Performances and Behavioral Interpretation

As the application of ML methods to predicting vehicle transactions is new, we first compare the performance among the four aforementioned ML models. **Figure 2** summarizes the overall performance evaluation on three well-known performance measures: overall accuracy, multi-class log loss and F1. Each point on the plot represents the average performance of a particular model over 5-fold cross-validation. For each ML method, models were generated based on several sets of features and therefore summarized with a boxplot. All the experiments were run using the MLJAR framework [46]. Results indicate that the two gradient boosting-based methods, CatBoost and LightGBM, are the best performing ML models. From the figure it can be seen that the CatBoost algorithm has smaller and tighter performance metric values than any other model tested here. Therefore, the CatBoost method is selected to further generate behavior interpretations.

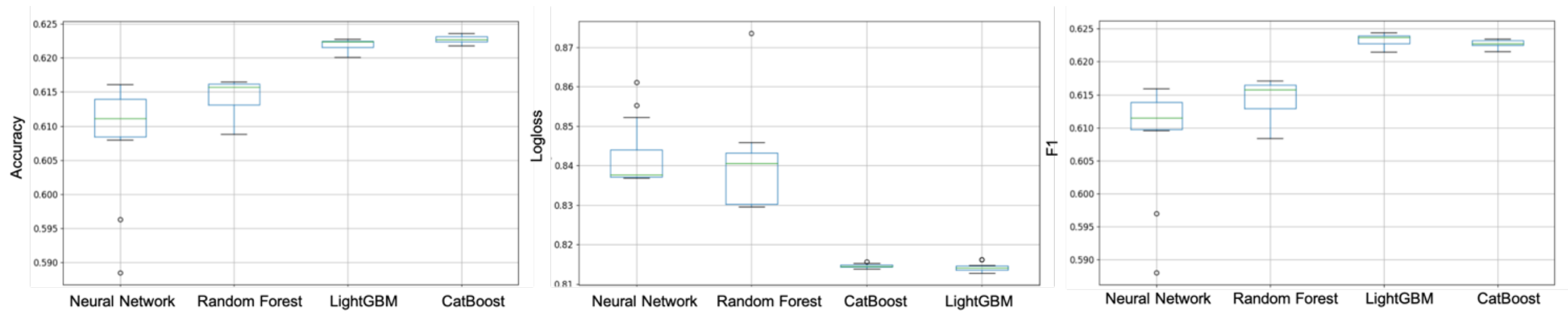

**Figure 2. Performance comparison for four machine learning models**

We use the local explanations (i.e., SHAP values) of individual input variables computed by the TreeExplainer to interpret effects of input variables on predicting the transaction outcome class ∈ {disposed, replaced, kept} for CatBoost predictions. SHAP values are computed for each observation of a vehicle's transaction outcome, with positive values indicating an increase in the outcome's log odds. The SHAP values are summarized with a set of beeswarm plots, where each dot corresponds to an observation in the dataset (**Figure 3)**. Each dot's position on the x-axis shows the impact a given variable has on the CatBoost model prediction for the vehicle's transaction outcome. When the multiple dots share the same effects (i.e., same x position), they are depicted as a swarm to indicate density.

Note that the dots are color coded by a variable's value from low (blue) to high (red) for the binary, ordinal, and continuous variables. Non-ordinal categorical variables such as vehicle type and race are colored in grey. The color spectrum reveals the direction (or lack thereof) of effects. For example, in **Figure 3(a)**, vintage is shown to have a negative association with a vehicle being "kept" in the household's fleet, which means that older vehicles (red dots corresponding to the vintage variable) are less likely to be kept in the next wave compared to younger ones. Similarly, vintage is somewhat positively correlated with being disposed or replaced i.e., older vehicles have a higher probability of being disposed or replaced compared to younger ones.

Variables are indicated on the y axis, ordered by the magnitude of their average SHAP value, indicating their global importance for predicting the respective outcome. The 25 most important variables are shown for each transaction outcome of interest.

**Figure 3(a)** summarizes the effect of each variable on whether a vehicle is less or more likely to be transacted out of the family (i.e., kept or not). **Figures 3(b)-(c)** further show the effect of the variables on the type of transaction (replacement or disposal). The variable importance ranking combining all outcomes are presented in **Figure 5 (a)**, which indicates that vehicle vintage and household fleet size are overall the most influential predictors.

**Figure 3(a)** shows that the vehicle-level attributes vintage and number of years serving the family (yrs_inFu) are the top predictors with opposite effects on whether or not a vehicle will be kept in the family. Household head or spouse' age (hh_age), education level (edu) and income level (inc_5bins) are the top sociodemographic attributes impacting vehicle transaction outcomes. Residential relocation, change in income, and change in number of drivers are among key life-events and change variables impacting vehicle transactions at the household-level.

When further differentiated by transaction types, **Figure 3(b)** and **3(c)** show different ordering of variable importance. While vintage is the leading predictor for replacement decisions pertaining to a vehicle, the household fleet size (Nveh) becomes the most important one for the disposal decision. Besides the importance ranking, the top 25 variables themselves are different in predicting replacement and disposal. Note that the presence of young children (kid_4 and kid_5_11) are important in predicting vehicle replacement but not disposal.

The direction of effects on transaction types are largely consistent for the vehicle attributes; any family attributes and life-events (e.g., Nveh, inc_5bins, ch_Ndriver, ch_income) have opposite effects (if the direction of effects can be discerned) on disposal vs replacement. For example, while larger fleet size (Nveh) increases the probability of disposal, it decreases the probability of replacement. Larger fleet size may lead to redundancy of vehicles and thus trigger vehicle disposal; however, a bigger fleet provides more diversity to satisfy various travel requirements and therefore could result in a lower need to replace a vehicle to fulfill any potential change in demand. Residential relocation increases the likelihood of disposal as well as replacement. The longer a vehicle has served the family (yrs_inFU), the less likely it is to be disposed or replaced. This is in the opposite direction of vehicle vintage, where older vehicles are more likely to be disposed or replaced.

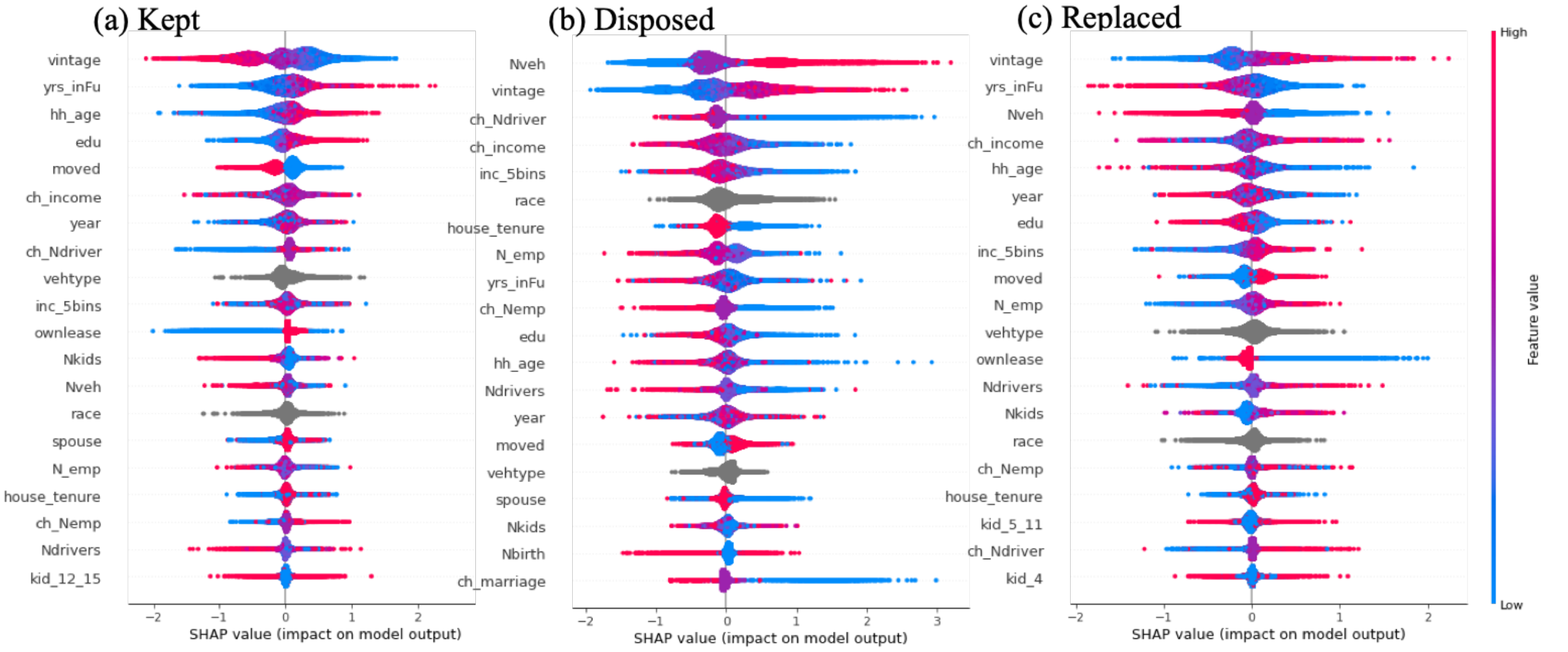

**Figure 3. Local explanation summary, i.e. variables' impact on the outcome class (a) kept, (b) disposed, and (c) replaced. The 25 most important input variables are shown for each outcome class and variables are ranked (from top to bottom) by their global importance measured by the average absolute SHAP value. See Table 1 for description of variable names.**

Besides the overall direction of effects, SHAP dependence plots are used to examine potential nonlinearity in the variable-to-outcome relationship for continuous variables such as vehicle vintage and change in income. **Figures 4 (a-b)** reveal that vehicle vintage has varying impacts on the likelihood of vehicle transaction outcomes of owned versus leased vehicles. The probability of 'keeping' a leased vehicle drops right around two to three years (coupled with an increase in probability of 'replacing' the vehicle) which is intuitive as the typical lease term is for 2-3 years. Starting from 11 years, vehicles, whether owned or leased, are expected to be transacted out of the family (i.e., SHAP value on "kept" becomes negative).

One interesting finding from the dependence plot **Figures 4 (a-b)** is that, once the leased vehicles are past the 3-year lease term period, they see a relatively 'quiet' period where increasing age (from 6-10 years) has no impact on the vehicle being replaced. This, however, does not hold true for owned vehicles, which see a steady or a stepped increase in the probability of getting replaced with increasing age. A plausible explanation for this phenomenon is households purchasing a leased vehicle once the lease term expires, if they are fully satisfied with the vehicle. Once the loan on the vehicle is paid off (2-3 years after the lease term, or 5-6 years after the original lease date), households would be reluctant to dispose or replace a vehicle that is fully paid off. This could, however, change as the vehicle ages (beyond 10 years), and maintenance costs and hassle outweigh the cost of replacement or disposal.

**Figures 4 (c-d)** reveal that there is generally a linear relationship between income change and vehicle disposal or replacement and this relationship differs for families of different income levels as indicated by the reversed vertical dispersions of the colors between positive and negative income changes. Such information provides data driven insights for binning the vintage variable and inclusion of interaction terms for the subsequent MNL analysis.

**Figures 4 (e-f)** indicate that the dependence of transaction behaviors on vehicle vintage also differs by household fleet sizes. Vehicles in single vehicle families are slightly less likely to be replaced early on than those in multi-vehicle families (Nveh >1) (**Figure 4 e)**. On the other hand, as they get older, vehicles in single vehicle families are more likely to be replaced than those in multi-vehicle families with extra vehicles to spare (**Figure 4 e)**. As for the disposal decision, vehicles in single vehicle families are less likely to be disposed as they get older (> 12 years) compared to those in families with extra vehicle(s) available.

The SHAP interaction scores are used to rank the importance of all possible combination of variables. Using the overall SHAP importance to select the top 20 features (**Figure 5a**), we further rank their interactions using SHAP interaction scores (**Figure 5b and 5c**). We find 5 out of 7 top interactions involve Nveh (household fleet size), indicating that the effects of other variables on the transaction decisions differ among families with different fleet sizes. The interaction scores also confirmed the importance of interactions between income levels and change in income as well as between vintage and owned versus leased, as visually evident in **Figure 4**.

Note that the age of the family head or spouse (hh_age) is a proxy for, and thus correlated with, many life events as well as family demographic status [47]. Although ML models can easily handle colinear features, collinearity issues may cause unstable estimates in logit regressions. As a result, age was not directly included as an independent variable in previous choice modeling (e.g., [10]). Instead, household age can be included as interaction terms with certain family demographic features. Here, **Figures 4 (g,h)** indicate that the presence of young children (<= 4 years and 5-11 years) increases the vehicle replacement probability in families of different parental age bins (20s and 30s), but not for later parental ages.

The feature importance ranking, as well as feature interactions, provide valuable data driven insights for variable selection, binning the vintage variable, and inclusion of interaction terms to improve the MNL model specification. The top-ranking variables and interactions identified by the SHAP importance ranking are shown to pass the LASSO testing (see Supporting Materials for more details) which further confirms their validity.

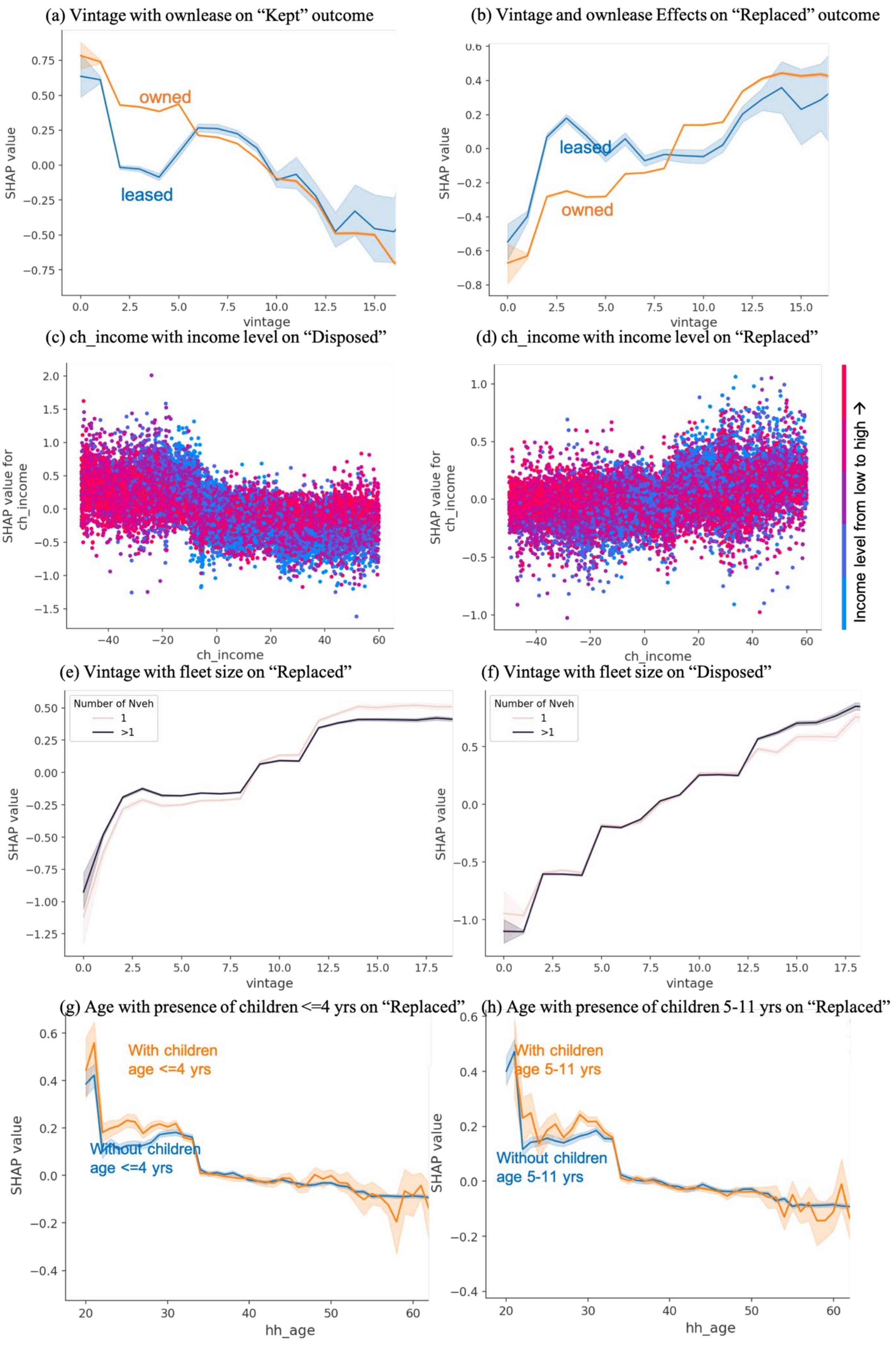
(a) Vintage with ownlease on "Kept" outcome
owned
leased
SHAP value
vintage
(b) Vintage and ownlease Effects on "Replaced" outcome
leased
owned
SHAP value
vintage
(c) ch_income with income level on "Disposed"
SHAP value for ch_income
ch_income
(d) ch_income with income level on "Replaced"
SHAP value for ch_income
ch_income
Income level from low to high →
(e) Vintage with fleet size on "Replaced"
Number of Nveh
1
>1
SHAP value
vintage
(f) Vintage with fleet size on "Disposed"
Number of Nveh
1
>1
SHAP value
vintage
(g) Age with presence of children <=4 yrs on "Replaced"
With children age <=4 yrs
Without children age <=4 yrs
SHAP value
hh_age
(h) Age with presence of children 5-11 yrs on "Replaced"
With children age 5-11 yrs
Without children age 5-11 yrs
SHAP value
hh_age

**Figure 4. SHAP dependence plots to illustrate effects of individual variables and their interactions on the transaction outcomes: effects of vehicle vintage and its interaction with the ownership type (owned or leased) on the likelihood of (a) "kept" outcome, and (b) "replaced" outcome; effects of change of income and its interaction with the income levels on the likelihood of (c) "disposed" outcome, and (d) "replaced" outcome; effects of vintage and its interaction with the household fleet size on the likelihood of (e) "replaced" outcome, and (f) "disposed" outcome; effects of household head or spouse age and its interaction with (g) presence of children of age <=4 years, and (h) presence of children of age 5-11 years, on "replaced" outcome.**

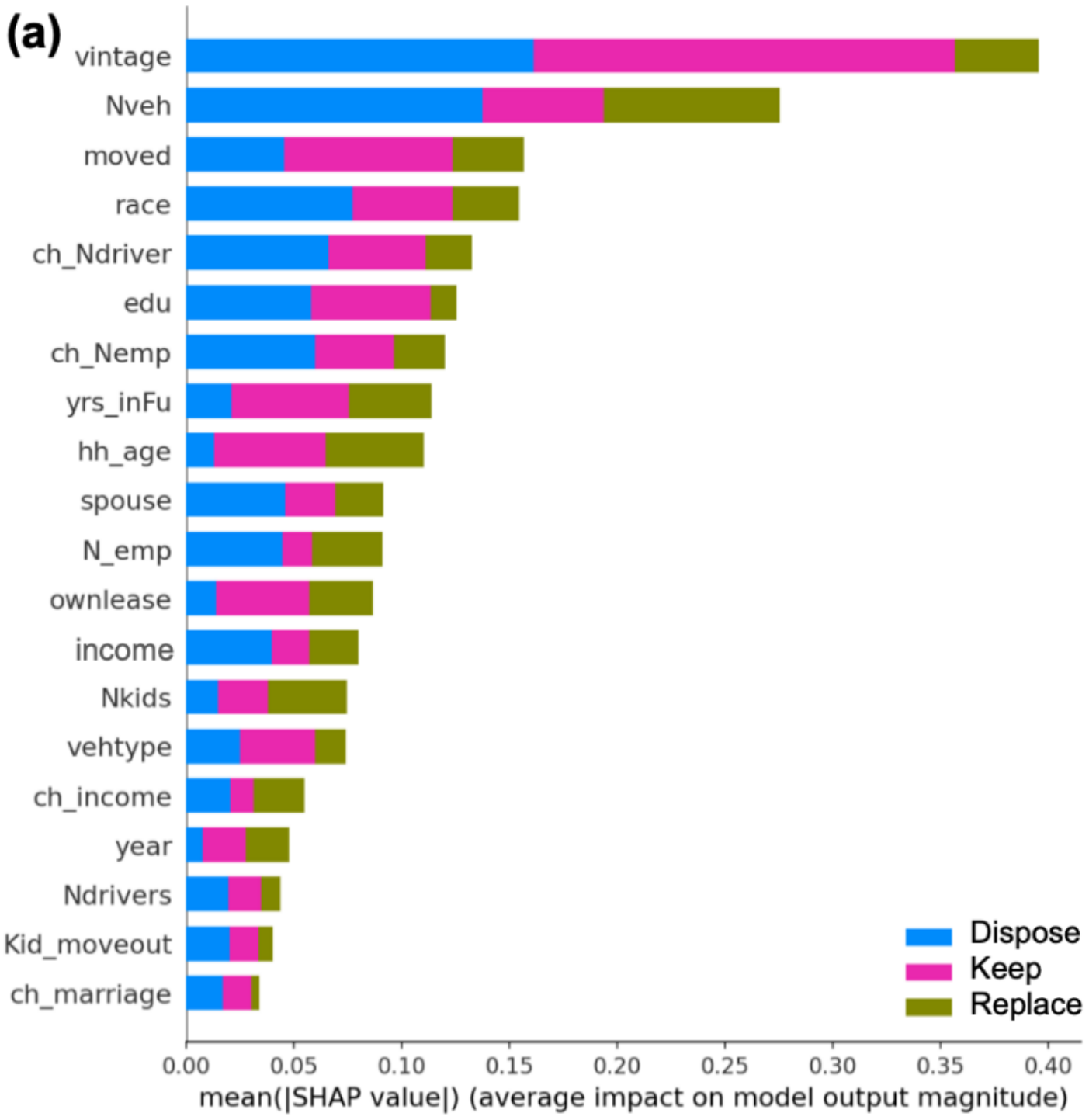

(b)

| Interaction Terms Predict "Replace" | SHAP Interaction Score |
|---|---|
| Nveh:Ndrivers | 13.9 |
| vintage:ownlease | 10.2 |
| Nveh:vintage | 7.7 |
| Nveh:inc_5bins | 7.0 |
| Nveh:Nkids | 6.9 |
| ch_income:income | 6.1 |
| Nveh:ch_income | 6.1 |

(c)

| Interaction Terms Predict "Dispose" | SHAP Interaction Score |
|---|---|
| Nveh:spouse | 55.8 |
| ch_Ndriver:Ndrivers | 36.1 |
| Nveh:income | 34.5 |
| ch_income:income | 31.9 |
| Nveh:race | 28.7 |
| Nveh:vintage | 26.4 |
| Nveh:ch_Ndriver | 23.1 |

**Figure 5. (a) SHAP Feature Importance considering all three outcomes; and interaction term scores for predicting "replaced" (b), and (c) "disposed".**

## 3.2 Multinomial logit Model Results and Interpretation

Informed by the SHAP values and dependence plots from the TreeExplainer, we generate vintage bins at < 2 yrs, 2-4 yrs, 5-10 yrs, and >=11 yrs to capture varying vehicle transaction propensities with vehicle vintage. We also include the interaction term of leased vehicles with vintage bin 2-4 yrs to further quantify the ownership type effects. Effects of income changes are interacted with income levels. We do not directly include the head of household age as a continuous independent variable; instead, we only include indicator variable of hh_age >60 to capture vehicle transactions due to driving cessation as household members get older [10]. Furthermore, informed by interactions between parental age with presence of children revealed by the TreeExplainer, we include parental age bins as an interaction term with the presence of young children to further distinguish the timing effects of these attributes. We separately estimate MNL models for the families with only one vehicle, and for families with extra vehicles (Nveh >1) to account for potential behavioral differences in transaction decisions due to the availability of vehicles as revealed in the TreeExplainer results. The estimation results are presented in **Table 2** and described below.

### 3.2.1 Fleet Size and Vehicle Attributes

Similar to the TreeExplainer results, fleet size is significantly associated with the probability of vehicle transactions with *opposite* effects: a larger household fleet increases the probability of disposal but decreases the probability of replacement. Effects of vehicle attributes differ between one-vehicle families and families with more than one vehicle. Despite older vehicles being more likely to be transacted out of the family, we find that one-vehicle families tend to hold on to vehicles a little longer before disposing them (indicated by the non-significant coefficient for vintage bin 2-4) compared to families with more than one vehicle. On the other hand, one-vehicle families are more likely to replace (rather than dispose of) their only vehicle, presumably in order to maintain their ability to address mobility needs.

The longer the vehicles stay with the family (yrs_inFu), the more likely they are to be kept. Such opposite effects between vintage and yrs_inFu was also evident in the SHAP results shown in **Figure 3**. This means that vehicles serving the family longer are less likely to be transacted out compared to ones that are of the same age but with shorter duration of ownership. This could be due to a sentimental attachment to the family vehicle or the longevity of the vehicles with less frequent switches of owners and better maintenance.

Owned vehicles are in general less likely to be transacted out of the household fleet compared to leased ones, especially for vehicles between 2 and 4 years old, which is consistent with the TreeExplainer results. The transaction likelihood also differs across body types. Light trucks such as pickups and SUVs are less likely to be disposed of or replaced compared to cars, while vans are more likely to be replaced in families with more than one vehicle. The reduced rate of disposal, particularly for pickup trucks, is corroborated by the increase in average age of a pickup truck from 10.1 years in 2011 to 13.1 years in 2017 [36].

### 3.2.2 Concurrent Family Status

As with the vehicle attributes, concurrent family socio-economic and demographic attributes are also found to impact vehicle transaction outcomes.

We observe that married (or cohabiting) families, families with higher education levels, home owners, and families with older parents tend to keep their vehicles longer. Note that while the literature generally do not include both number of eligible drivers and the number of works in the model, we include them together as they are both selected by the SHAP importance ranking. Families with more workers are less likely to dispose of their vehicles, presumably owing to the necessity of individual household members requiring their own vehicle to commute to work (note that the data used for modeling was from pre-pandemic times). Families with more driving age members are more likely to dispose of or replace their vehicles, which could stem from variation in taste across different drivers, or different requirements for different drivers in the family. The result of higher number of drivers associated with higher likelihood of disposing vehicles may seem surprising, although we note that the inclusion of number of workers and income as control variables may help explain this: the positive coefficient of number of eligible drivers on disposal decision should be interpreted as "the effect of more drivers while number of workers and income remains the same".

There is noticeable variation in transaction probabilities of white families with more than one vehicle, who are less likely to dispose and more likely to replace their vehicles compared to their Asian counterparts. Note that this finding indicates a potential culture difference among different racial communities or a confounding effect through our omission of location factors (such as residential density) that correlated with race [48]. Transaction probability also varies among income groups. Poorer families are associated with a higher disposal probability and lower replacement probability than more affluent families, possibly due to lack of financial stability to maintain a vehicle.

Presence of young children (kid_4 and kid_5_11) was found to mostly affect replacement decisions in the TreeExplainer results. The MNL model results confirm this and further distinguish timing differences. Interestingly, we find that parents of age 35 or younger are more likely to replace their vehicles when young children (<=11 years old) are present. In contrast, the replacement decisions for parents older than 35 are not associated with the presence of young children. This finding is consistent with previous literature that has found that travel behaviors are more frequently changed before age 35 years old [21] .

### 3.2.3 Life Events and Change Variables

Life events generally change mobility needs of families, which in turn necessitate vehicle transactions. A number of events, such as residential relocation, grownup children moving out of the family (i.e., empty nesting), decoupling, and childbirth are found to increase vehicle transaction (disposal or replacement) probabilities. On the other hand, increasing level of education, increasing number of workers or drivers in the family, and coupling tend to decrease the vehicle disposal probability.

The interaction effects between income change and income groups observed in the TreeExplainer are also confirmed in the MNL results. An increase in family income decreases the probability of vehicle disposal, while it increases the probability of vehicle replacement. This effect is more significant for poorer families. The transaction decisions of the top earners, on the other hand, are unaffected by the income changes.

While it is intuitive that a childbirth event increases the probability for vehicle replacement (presence of children is strongly associated with owning larger vehicles [49]), our analysis reveals that the effects are most significant among younger parents (< 27 years old). The birth event for these parents is more likely to be their first child, and their existing family fleet is likely less tailored to meet parenting needs.

**TABLE 2 MNL Estimates (the reference level of vehicle outcome is "kept").**

| **Explanatory Variables** | **Families with 1 Veh** | | **Families with >1 Veh** | |
|---|---|---|---|---|
| | Disposed | Replaced | Disposed | Replaced |
| Constant | **-1.887*** ** | **-0.601*** | **-2.894*** ** | **-0.568*** ** |
| | (0.463) | (0.256) | (0.253) | (0.16) |
| ***Current Fleet Size*** | | | | |
| Number of vehicles in family unit | | | **0.802*** ** | **-0.070*** ** |
| | | | (0.029) | (0.02) |
| ***Vehicle Attributes*** | | | | |
| Vehicle vintage (reference level is <2 yrs) | | | | |
| 2-4 yrs | 0.098 | **0.354** ** | **0.538*** ** | **0.255*** ** |
| | (0.244) | (0.127) | (0.126) | (0.064) |
| 5-10 yrs | **0.529*** | **0.725*** ** | **1.037*** ** | **0.548*** ** |
| | (0.24) | (0.125) | (0.126) | (0.065) |
| >= 11 yrs | **1.248*** ** | **1.416*** ** | **1.779*** ** | **1.093*** ** |
| | (0.241) | (0.129) | (0.129) | (0.068) |
| Number of years serving the family | **-0.050*** ** | **-0.060*** ** | **-0.032*** ** | **-0.053*** ** |
| | (0.01) | (0.006) | (0.005) | (0.004) |
| Vehicle is owned (rather than leased) | **-0.392** ** | **-0.408*** ** | **-0.445*** ** | **-0.517*** ** |
| | (0.138) | (0.093) | (0.127) | (0.079) |
| Leased * Vintage in 2-4 yrs | **0.702**** | **0.885***** | **0.766***** | **1.260***** |
| | (0.241) | (0.144) | (0.176) | (0.102) |
| Body type (reference is car body type) | | | | |
| Pickup | **-0.386**** | **-0.152*** | **-0.229***** | **-0.242***** |
| | (0.136) | (0.077) | (0.049) | (0.03) |
| Utility | **-0.238**** | **-0.159***** | **-0.141**** | **-0.106***** |
| | (0.081) | (0.046) | (0.043) | (0.026) |
| Van | -0.055 | 0.083 | -0.018 | 0.106** |
| | (0.124) | (0.078) | (0.066) | (0.04) |
| ***Concurrent Family Attributes*** | | | | |
| Married or cohabiting | -0.051 | 0.019 | **-0.862***** | **-0.083*** |
| | (0.098) | (0.06) | (0.051) | (0.039) |
| Income levels (reference level is $50,000 – $75,000) | | | | |
| < $25,000 | **0.560***** | **-0.278***** | **0.341***** | -0.048 |
| | (0.122) | (0.066) | (0.099) | (0.081) |
| $25,000 – $50,000 | **0.257*** | -0.08 | **0.146*** | **-0.133**** |
| | (0.109) | (0.053) | (0.065) | (0.048) |
| $75,000 - $150,000 | -0.125 | 0.088 | **-0.118*** | **0.086**** |
| | (0.171) | (0.062) | (0.051) | (0.031) |
| >= $150,000 | 0.406 | 0.113 | 0.055 | **0.210***** |
| | (0.245) | (0.099) | (0.065) | (0.04) |
| Race (reference is Asian) | | | | |
| White | -0.008 | -0.016 | **-0.291*** | **0.181*** |
| | (0.331) | (0.177) | (0.141) | (0.09) |
| Other race categories not significant and not shown | | | | |
| Education level (reference is less than high school) | | | | |
| Post graduate | **-1.081***** | **-0.220*** | **-0.356***** | **-0.302***** |
| | (0.166) | (0.094) | (0.098) | (0.071) |
| College | **-0.922***** | **-0.271**** | **-0.192*** | **-0.248***** |
| | (0.141) | (0.088) | (0.096) | (0.069) |
| Some college | -0.489*** | -0.053 | -0.174 | -0.138* |
| | (0.104) | (0.078) | (0.09) | (0.066) |
| High school | -0.332*** | -0.043 | -0.052 | -0.051 |
| | (0.098) | (0.078) | (0.088) | (0.066) |
| Number of eligible drivers | **0.324***** | **0.115**** | 0.019 | **0.072***** |
| | (0.062) | (0.04) | (0.034) | (0.021) |
| Number of workers | **-0.619***** | 0.015 | **-0.208***** | **0.087***** |
| | (0.071) | (0.042) | (0.032) | (0.021) |
| Head or spouse age > 60 yrs old | **-0.194*** | **-0.311***** | -0.043 | **-0.104**** |
| | (0.097) | (0.062) | (0.055) | (0.036) |
| Number of children (<16 yrs) | 0.058 | 0.022 | 0.061 | **0.066**** |
| | (0.078) | (0.05) | (0.044) | (0.024) |
| Presence of children <=4 yrs old for | | | | |
| Parents <27 yrs old | 0.315 | **0.455***** | -0.103 | 0.16 |
| | (0.173) | (0.125) | (0.148) | (0.1) |
| Parents 27 – 35 yrs old | -0.253 | -0.071 | -0.065 | 0.052 |
| | (0.162) | (0.1) | (0.09) | (0.051) |
| Parents > 35 yrs old | -0.183 | 0.14 | -0.089 | -0.07 |
| | (0.196) | (0.117) | (0.087) | (0.051) |
| Presence of children 5-11 yrs old for | | | | |
| Parents <27 yrs old | -0.116 | 0.207 | 0.205 | 0.131 |
| | (0.198) | (0.146) | (0.203) | (0.143) |
| Parents 27 – 35 yrs old | -0.089 | **0.247*** | 0.034 | **0.117*** |
| | (0.158) | (0.097) | (0.097) | (0.055) |
| Parents > 35 yrs old | -0.273 | 0.025 | -0.112 | -0.006 |
| | -0.116 | 0.207 | 0.205 | 0.131 |
| Presence of children 12-15 yrs old | 0.022 | -0.005 | 0.108 | -0.009 |
| | (0.131) | (0.08) | (0.071) | (0.043) |
| Is home owner | **-0.434*** ** | -0.065 | **-0.265*** ** | -0.053 |
| | (0.093) | (0.051) | (0.05) | (0.034) |
| ***Life events and change variables*** | | | | |
| Change in head marriage (coupling up = 1, no change = 0, decoupling = -1) | **-0.686*** ** | -0.117 | **-1.366*** ** | -0.098 |
| | (0.188) | (0.089) | (0.085) | (0.065) |
| Change in income (in thousands) for families of income levels: | | | | |
| < $25,000 | **-0.012** ** | **0.005** ** | -0.005 | **0.004*** |
| | (0.004) | (0.002) | (0.003) | (0.002) |
| $25,000 – $50,000 | **-0.010** ** | 0.002 | **-0.005** ** | **0.002*** |
| | (0.004) | (0.001) | (0.002) | (0.001) |
| $50,000 – $75,000 | -0.008 | 0.001 | **-0.004*** | **0.002** ** |
| | (0.004) | (0.001) | (0.002) | (0.001) |
| $75,000 - $150,000 | -0.007 | 0.001 | **-0.003*** ** | **0.001** ** |
| | (0.004) | (0.001) | (0.001) | (0.0003) |
| >= $150,000 | 0.0001 | 0.0003 | 0.0001 | -0.00001 |
| | (0.0005) | (0.0003) | (0.0002) | (0.0001) |
| Increase level of education | -0.293 | 0.065 | **-0.197*** | 0.006 |
| | (0.161) | (0.086) | (0.083) | (0.049) |
| Birth of children for | | | | |
| Parents <27 yrs old | 0.268 | **0.349** ** | 0.248 | **0.323** ** |
| | (0.175) | (0.13) | (0.155) | (0.099) |
| Parents 27 – 35 yrs old | 0.018 | 0.179 | -0.155 | 0.085 |
| | (0.186) | (0.101) | (0.085) | (0.046) |
| Parents > 35 yrs old | 0.295 | -0.251 | 0.206 | 0.033 |
| | (0.22) | (0.146) | (0.109) | (0.062) |
| Empty nest (grown up children move out) | 0.095 | **0.351** ** | **0.370*** ** | 0.061 |
| | (0.178) | (0.121) | (0.069) | (0.052) |
| Change in number of workers | **-0.619*** ** | -0.047 | **-0.298*** ** | 0.029 |
| | (0.066) | (0.038) | (0.034) | (0.021) |
| Change in number of adults | **-0.203*** | **0.119** ** | **-0.324*** ** | -0.004 |
| | (0.08) | (0.045) | (0.045) | (0.028) |
| Recently moved | **0.569*** ** | **0.337*** ** | **0.382*** ** | **0.284*** ** |
| | (0.08) | (0.043) | (0.039) | (0.026) |
| Controlled for year fixed effects | Yes | | Yes | |
| Observations | 16,510 | | 52,390 | |
| Log Likelihood | -13,631 | | -43,078 | |
| LR Test | 2,874.534*** (df = 122) | | 8,420.184*** (df = 124) | |

Note: *p<0.05; **p<0.01; ***p<0.001

## 3.3 Performance Comparison between ML and MNL Models

The performances of the resulting improved MNL model (iMNL), as well as the Baseline MNL (bMNL) and the best performing ML model are evaluated and compared using the 10 metrics presented in **Table 3,** for both in-sample data (i.e., training sample) and testing data (i.e., out-sample data from a random selection of 1000 households).

Although ML model performance is better than the Baseline MNL model, for both in-sample and out-of-sample, the differences are more pronounced with in-sample and diminish once both models are evaluated on testing data.

Similar to the Baseline MNL, the improved MNL model shows poorer performance than the ML model on in-sample data. However, when applied to the testing data, 5 out of the 10 metrics have now indicated same or better performance of the improved MNL model compared to the ML model, and the performance differences are smaller between the improved MNL and ML compared to between the Baseline MNL and ML models. Furthermore, **Table 3** suggests overall MNL models perform more consistently between in-sample and out-of-sample data than ML models.

**TABLE 3 Performance metrics for baseline multinomial logit (bMNL), improved multinomial logit (iMNL), and the best performing machine learning (bML). Best performing metrics are indicated with bold faces.**

| Metrics | In-sample | | | Testing Sample | | |
|---|---|---|---|---|---|---|
| | bMNL | iMNL | bML | bMNL | iMNL | bML |
| Overall Accuracy | 0.61 | 0.62 | **0.72** | 0.61 | **0.62** | **0.62** |
| Average Accuracy | 0.74 | 0.74 | **0.81** | 0.74 | **0.75** | **0.75** |
| Macro-precision | 0.53 | 0.53 | **0.75** | 0.53 | **0.55** | 0.53 |
| Sensitivity | 0.42 | 0.42 | **0.58** | 0.42 | 0.43 | **0.44** |
| Macro-F1 | 0.41 | 0.42 | **0.62** | 0.42 | 0.43 | **0.45** |
| Micro metrics | 0.61 | 0.62 | **0.72** | 0.61 | **0.62** | **0.62** |
| Cohen's Kappa | 0.16 | 0.17 | **0.42** | 0.17 | 0.19 | **0.21** |
| Specificity | 0.71 | 0.72 | **0.79** | 0.72 | 0.72 | **0.73** |
| Cross Entropy | 1.59 | 1.57 | **1.44** | 1.58 | 1.58 | **1.47** |
| 1/(Log Loss) | 1.20 | 1.22 | **1.52** | 1.20 | **1.22** | 1.20 |

# 4 Discussions and Conclusions

## 4.1 Contribution to Travel Behavior Literature

In this paper, we find the dynamic decisions to let go of a given vehicle (through disposal or replacement) are positively correlated with (1) the age of the vehicle coupled with the vehicle being leased rather than owned; (2) demographic characteristics such as families with a greater number of drivers, and/or lower income level; and (3) key life cycle events such as childbirth (particularly for younger parents), residential relocation and empty nesting. On the other hand, factors positively associated with households' decision of holding on to older vehicles are (1) vehicle attributes, such as that a vehicle is owned rather than leased, and that the vehicle is a pickup truck or an SUV (as opposed to a passenger car); (2) demographic attributes such as having a higher level of education or being a homeowner; and (3) life events such as marriage or reduction of family income. While past studies have separately investigated one or two of the above dimensions, this paper is the first to include all three simultaneously based on revealed preferences in the PSID panel survey, for both vehicle attributes and household characteristics.

Furthermore, our empirical analysis methodology provides innovation beyond previous literature by leveraging machine learning coupled with TreeExplainer as an additional interpretation tool to both generate behavioral insights and improve the model specification for MNL choice modeling. This study represents the first application of ML methods to model vehicle transactions using a large panel dataset. We find the two gradient-boosting-based methods, CatBoost and LightGBM, are the best performing ML models for this problem. We demostrate that using SHAP interpretation tools coupled with multinomial logistic models could help them achieve similar performance levels to the best performing ML methods. The variable effects, in terms of the direction of influence, are largely consistent between the two methods, which improve the robustness of the behavior insights generated by our study.

## 4.2 Policy Implications

Effects of vehicle attributes estimated from this study have several policy implications, particularly in the context of increased policy interest in accelerating turnover of the vehicle fleet in the aid of transportation decarbonization. While older vehicles are more likely to be transacted out of the family than newer ones, the transaction probability decreases as the vehicles serve the family longer. A similar pattern was observed in a previous stated-preference vehicle survey [49]. Both the temporal trend in our data (**Figure 1 b**) and the NHTS from 2009 to 2017 [36] have revealed a national increase in average vehicle age in U.S. households. In essence, the longer a family holds on to a vehicle, the more likely the household will continue to retain that vehicle. This is an important pattern to identify and further understand, as it provides a positive feedback loop to lengthen the vehicle replacement schedule and consequently slow down the penetration of emerging, and potentially preferred technologies from a policy perspective. Policies such as vehicle replacement or disposal incentives (such as bounties to retire older vehicles) could be introduced to help break this feedback loop. In future research we will analyze the potential effectiveness of alternative policies to accelerate vehicle fleet turnover using our model.

Transaction probabilities also differ by vehicle body type, with light trucks, such as pickup trucks and SUVs, less likely to be replaced or disposed of than cars. This effect is consistent with the stated preference survey results in [49]. Light trucks are the most popular body types in states throughout the U.S. other than California [50], with their market share in light duty vehicle sales reaching 72% in 2019 and 77% in 2020 [15]. Currently, electric vehicles (EVs) are mostly available for cars rather than light trucks [50]. Without a more diverse supply of electric light trucks, the increasing ownership of conventional light trucks coupled with their low turnover rates may slow down the overall penetration of EVs.

Another important but often omitted vehicle attribute in previous studies is ownership status (i.e., whether a vehicle is owned or leased). We find that leased vehicles are much more likely to be replaced or disposed of relative to owned vehicles, with the difference greatest at 3 years of age, consistent with the typical leasing term. Leased vehicles now represent 31% of all new car transactions, a rate that has increased more than 10% per year since 2014 [51]. Finite lease terms contribute to accelerating adoption of emerging vehicles; it is therefore important to understand trends in the lease market in order to forecast long-term car ownership trends and penetration of new vehicle technologies such as electric vehicles.

## 4.3 Contribution to Mobility Biography Literature

Following the mobility biography approach, our study has also examined the mediating and triggering effects of family socio-demographic status and life events on vehicle transactions. We find that married families, families with higher education levels, home owners, and families with older heads of household tend to keep their vehicles longer. Life events such as child birth, residential relocation, and change of household composition and income are found to increase one or both types of transactions (disposal or replacement), with signs largely consistent with literature findings [2], [13]. We further find that the poor families are more sensitive to income changes than more affluent families in the context of vehicle transactions.

Note that the two types of vehicle transaction we examined in this study represent different changes in the vehicle holdings at the family level. Vehicle disposal leads to a decrease in the level of vehicle holdings, while a replacement outcome maintains the overall fleet size of the household. Therefore, these two transaction types are expected to have different sensitivity to family level attributes and life events. Indeed, this is a pattern we confirm in our results. In particular, presence of children and child birth events are only significantly associated with the probability of vehicle replacement, not disposal. Furthermore, such effects strongly depend on parental age; in general, younger parents are more likely to replace their vehicles upon entering parenthood or with young children at home. One potential policy implication of this insight, in the context of policies designed to more rapidly turn over the vehicle fleet, is to design emerging vehicles with desired attributes for targeting different consumer segments.

### 4.4 Future Directions

While this study presents a comprehensive picture of the impacts of vehicle attributes, family demographics, and life events on vehicle transactions, it does have limitations that need further research. First, information about vehicle powertrain technology (such as hybrid, battery electric, or conventional internal combustion) was collected only for the more recent survey waves and thus was not included in the vehicle attributes in this analysis. Powertrain information will be available in future survey waves, which will enable this information to be included as an independent variable to understand the differences in disposal/replacement tendencies between vehicles powered by emerging technologies and conventional ones. Second, we have not yet included location factors (such as accessibility to alternative or shared travel modes) nor neighborhood characteristics (such as population density, job density, etc.) [10]. Third, as life events are determined from the two-year window between waves, broader lead or lag effects (see [14]) may be included to understand the more enduring effects associated with certain life events such as the duration of marriage [21]. Lastly, the present study only seeks to understand how soon a given vehicle will be transacted out of the household fleet. In order for emerging vehicle technologies to penetrate further into the household fleet, the next questions to address are what type of vehicles will be added to the household upon the replacement event, and how the vehicle choices are associated with household characteristics, and how they could be potentially influenced by new vehicle attributes and policy levers. In future work we will integrate a vehicle/technology choice model with a vehicle transaction model in order to assess the effectiveness of alternative policies to changing over the vehicle fleet, facilitate rapid penetration of desired technologies for transportation decarbonization, or otherwise influence vehicle transaction and vehicle choice behavior.

While this study demonstrates an innovative use of ML methods to inform the choice model building process, additional investigation and advancement of the methodology can be made. First, methodology needs to be developed to translate the local interpretation by SHAP values to a more straightforward global elasticity quantification similar to the MNL coefficients in a multinomial context. Second, this study focuses on tree-based methods while deep learning methodologies, such as [52], [53], can also be used to generate behavior interpretation. Future work could apply these deep learning models to our dataset and compare the performance and interpretation. Lastly, the data-driven insights derived from ML models can be applied to improve alternative modeling framework such as continuous time-to-the event models.

## Author Contributions

The authors confirm contribution to the paper as follows: principal investigator: C. Anna Spurlock; lead of the study design and manuscript writing: Ling Jin; data analysis: Ling Jin, Alina Lazar, Caitlin Brown, Qianmiao Chen, and Tin Ho; contributing to manuscript writing: C. Anna Spurlock, Venu Garikapati, Binrong Su, Srinath Ravulaparthy, Alex Sim; data and method supervision: Kesheng Wu and Tom Wenzel; and manuscript editing: all coauthors. All authors reviewed the results and approved the final version of the manuscript.

## Funding

This paper and the work described were sponsored by the U.S. Department of Energy (DOE) Vehicle Technologies Office (VTO) under the Systems and Modeling for Accelerated Research in Transportation (SMART) Mobility Laboratory Consortium, an initiative of the Energy Efficient Mobility Systems (EEMS) Program, under Lawrence Berkeley National Laboratory Contract No. DE-AC02-05CH11231.

## Disclosure

Parts of the results were presented in IEEE Bigdata 2021.